\documentclass[12pt]{iopart}
\usepackage{soul}
\usepackage{color}
\usepackage{todonotes}
\usepackage[colorlinks=true,citecolor=blue]{hyperref}
\usepackage[]{graphicx}
\usepackage{subcaption}
\newcommand{\argmin}{\mathop{\mathrm{arg\,min}}}
\newcommand{\argmax}{\mathop{\mathrm{arg\,max}}}
\newcommand{\Hat}[1]{\expandafter\hat#1}  
\begin{document}

\title[]{Deep learning for Gaussian process tomography model selection using the ASDEX Upgrade SXR system}

\author{F Matos$^1$, J Svensson$^2$, A Pavone$^2$, T Odstr\v{c}il$^3$, F Jenko$^1$ and the ASDEX Upgrade Team\footnote[1]{See author list of H. Meyer et al. 2019 Nucl. Fusion 59 112014} }

\address{$^1$ Max Planck Institute for Plasma Physics, Boltzmannstr. 2, 85748 Garching, Germany}
\address{$^2$ Max Planck Institute for Plasma Physics, Wendelsteinstr. 1, 17491 Greifswald, Germany}
\address{$^3$Plasma Science and Fusion Center, Massachusetts Institute of Technology, Cambridge, MA, USA}

\ead{francisco.matos@ipp.mpg.de}
\vspace{10pt}

\begin{abstract}
Gaussian process tomography (GPT) is a method used for obtaining real-time tomographic reconstructions of the plasma emissivity profile in a tokamak, given some model for the underlying physical processes involved. GPT can also be used, thanks to Bayesian formalism, to perform model selection -- i.e., comparing different models and choosing the one with maximum evidence. However, the computations involved in this particular step may become slow for data with high dimensionality, especially when comparing the evidence for many different models. Using measurements collected by the ASDEX Upgrade Soft X-ray (SXR) diagnostic, we train a convolutional neural network (CNN) to map SXR tomographic projections to the corresponding GPT model whose evidence is highest. We then compare the network's results, and the time required to calculate them, with those obtained through analytical Bayesian formalism. In addition, we use the network's classifications to produce tomographic reconstructions of the plasma emissivity profile, whose quality we evaluate by comparing their projection into measurement space with the existing measurements themselves. 

\end{abstract}

\vspace{2pc}

\section{Introduction}


Computed tomography generally refers to the process of imaging the interior of a body through indirect measurements. In many applications, this is achieved by focusing penetrating radiation on an object of interest from several directions and measuring the resulting decrease in radiation intensity on the opposite side (due to absorption by the body itself). Use of this information, the so-called \textit{projection} of the object, allows one to reconstruct its internal properties\cite{kak88principles}. 

In the case of radiative bodies, an alternative way to determine their properties is to perform cross-sectional imaging by treating the emitted radiation itself as a projection\cite{ingesson2008chapter}. In the field of nuclear fusion, this procedure is employed in many tokamaks for the reconstruction of plasma emissivity profiles\cite{mlynar2010inversion}. More specifically, in the ASDEX Upgrade tokamak, such imaging can be done with information from the soft X-Ray (SXR) diagnostic, which measures the line-integrated radiation emitted by the plasma along several lines of sight; these can be used to perform tomographic reconstruction (or inversion) of the plasma emissivity profile. Knowledge of this is useful for exploring magnetohydrodynamic phenomena, in addition to studying accumulation of impurities inside the plasma (particularly tungsten) due to their large contribution to the total amount of radiation\cite{odstrcil2016optimized}. 


Several techniques exist for solving the tomography problem\cite{mlynar2019current}. One approach is to use regularization-based algorithms, namely Tikhonov- and minimum Fisher-based techniques\cite{loffelmann2016minimum}. More recently, work has also been done using machine learning methods, namely deep neural networks\cite{ferreira2019deep, matos2017deep, jardin2019neural}, that are trained to create new reconstructions based on existing ones. 

Yet another method is Gaussian process tomography (GPT)\cite{svensson2011non}. GPT is an established method for performing tomographic inversion on many different types of physical distributions, that are modeled as posterior Gaussian distributions in a Bayesian setting. Computing a posterior first requires specifying a prior distribution, which encodes one's assumptions about the underlying physical process before any measurements of it are taken. The posterior can then be computed based on that prior, and on an observation (measurement) of the data generated by the physical process. The prior itself can either be a fixed distribution, or be drawn from a family of different models. 

Knowing the posterior, GPT guarantees that one can obtain the most likely (maximum \textit{a posteriori}, MAP) estimate for the tomographic reconstruction as well as its associated error values. More interestingly, however, through Bayesian inference, GPT prescribes a way to estimate the evidence for different models, through a process known as Bayesian model selection. This procedure can be of particular importance in cases where the choice of prior might have a strong effect on the results of the tomographic inversion. 

Unfortunately, in a neural network, there are no guarantees\cite{blundell2015weight} about whether the reconstructions obtained correspond to the MAP estimate of the underlying distribution, and there is no direct way, in standard Deep Neural Networks, such as convolutional neural networks, to obtain uncertainty estimates on the outputs. Bayesian neural networks\cite{gal2016dropout, pavone2018bayesian} and generative adversarial networks (GANs)\cite{radford2015unsupervised} can generate probability distributions for their outputs; however, they can be computationally expensive and, in the case of GANs, difficult to train\cite{salimans2016improved}.  

On the other hand, neural networks essentially store whatever function they have learned (through their training process) in their weights, making the inference process for new data very fast. With GPT, computing the MAP estimate based on a fixed model is also sufficiently fast for real-time purposes. This does not necessarily hold true, however, when performing bayesian model selection, since the process requires a series of additional computational steps, namely matrix inversions or using non-linear optimizers, which can be time-consuming, especially for data with a high dimensionality. 



Thus, we propose an approach where we train a convolutional neural network (CNN) to learn the GPT model selection procedure. To do this, we take SXR measurement samples from several ASDEX Upgrade shots and, through Bayesian model selection formalism, compute for each data point the corresponding model (out of a set of possible, pre-defined ones) with the highest evidence. We then train the CNN to reproduce this step, i.e., to map measurements to their highest evidence model. Finally, through the GPT framework, we compute the tomographic reconstruction of the plasma emissivity profile for each measurement, given the most likely models predicted by the CNN. 


This paper is organized as follows. Section \ref{sec:background} gives an overview of the problem of tomography, in particular soft x-ray tomography ASDEX Upgrade tokamak, and the existing techniques to solve it, including a review of GPT with bayesian model selection. Section \ref{sec:methods} details the data we collected, the formulation of our problem, and the model proposed to solve it. Section \ref{sec:results} details the direct results of the neural network classification, and the tomographic reconstructions obtained based on them; section \ref{sec:conclusions} describes and discusses our conclusions.

\section{Background}
\label{sec:background}

\subsection{Computed Tomography}

\begin{figure*}[h!]
    \centering
        \includegraphics[scale=0.5, trim=0 0 0 0, clip]{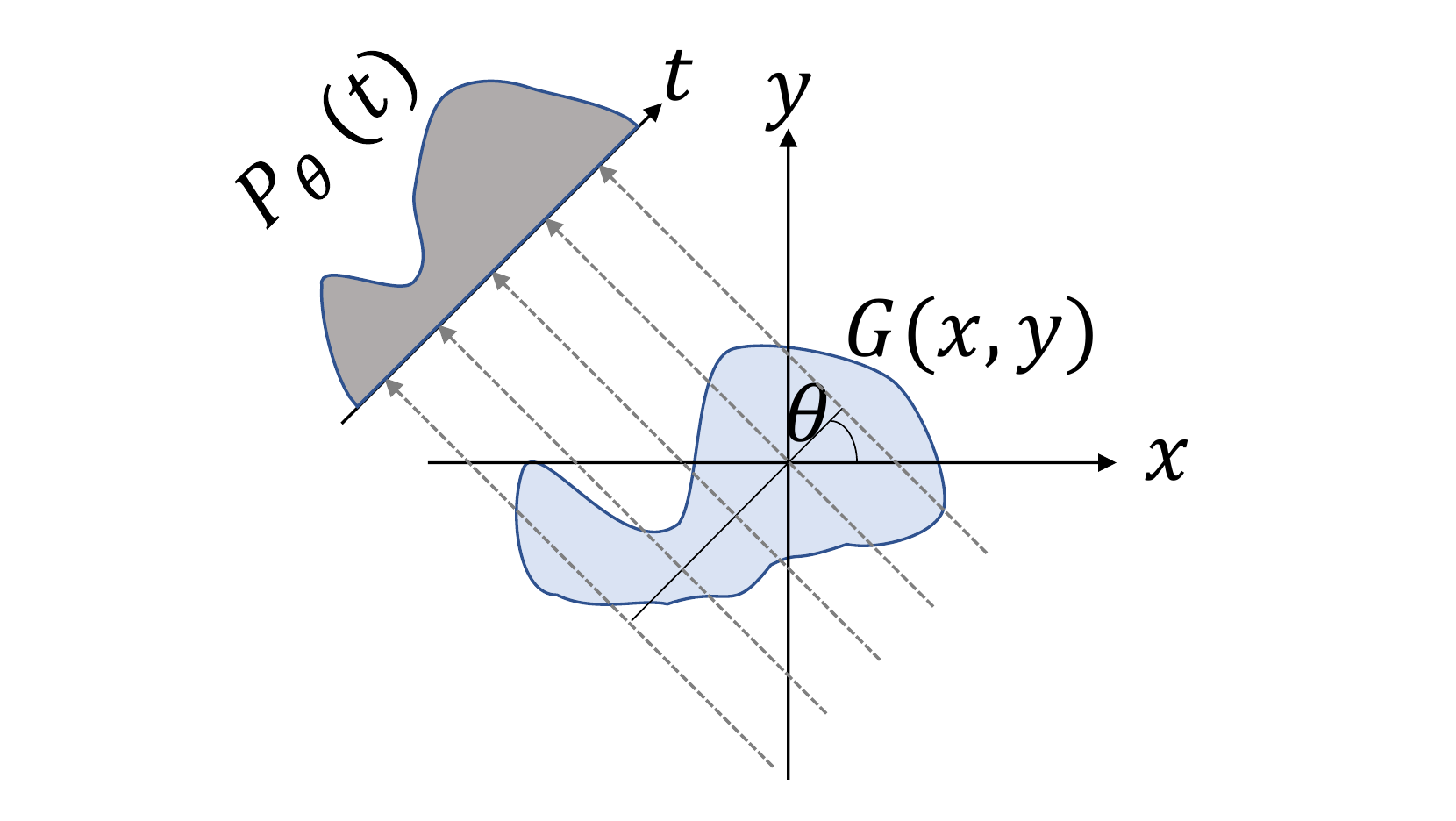}
        \caption{An illustrated projection, $P_{\theta}$, measured along an angle $\theta$. The blue area $G(x,y)$ is the cross section of interest, and is being traversed by radiation. Because different rays traverse different areas of the object, the value at each point $t$ in the projection space will be different.}
        \label{fig:projection}
\end{figure*}

The purpose of tomography is to reconstruct the internal (either two- or three-dimensional) properties of a given body from non-local measurements. Traditional tomographic algorithms achieve this by acquiring several \textit{projections} of the object of interest from different (potentially all) directions\cite{kak88principles}. Mathematically, a projection is a function that computes the line-integrated absorbency (or, in the case of fusion plasmas, emissivity) of a body along several paths or lines of sight (LOSs) as

\begin{equation}
    P_{\theta}(t) = \int\displaylimits_{L(\theta, t)} G(x,y) \,dL
\end{equation}{}

where $t$ is a point in the projection domain, $L(\theta, t)$ is the LOS crossing the body mapping to $t$ (along a direction given by an angle $\theta$), and $G(x,y)$ is the two-dimensional physical distribution of interest (see Figure \ref{fig:projection}). 

By computing several tomographic projections with different directions (i.e. different values of $\theta$), it is possible to reconstruct $G(x,y)$. For an exact reconstruction based only on the projections, an infinite number of them would need to be obtained. In practice, in many settings such as nuclear fusion experiments, it is difficult, or impossible, to obtain more than a handful of such projections, and some additional information, in the form of assumptions about the function $G(x,y)$, must be introduced in order to obtain a tomographic reconstruction. 


\subsection{SXR tomography at ASDEX Upgrade}

In the ASDEX Upgrade Tokamak, the Soft X-ray (SXR) diagnostic\cite{igochine2010hotlink} consists of eight pinhole cameras that measure the total radiation emitted by the plasma along 208 different volumes of sight (VOSs)\cite{odstrcil2016optimized}. We considered the extent of the VOSs in the toroidal and poloidal directions of the tokamak to be minimal, and treated them instead as lines of sight (LOSs). In addition, we also ignored the fact that the LOSs in the same camera array partially overlap. Based on this, the measurements collected by the individual cameras correspond to a single projection of the underlying plasma emissivity distribution, which is computed at 208 discrete positions, in a poloidal plane. In terms of the poloidal coordinates $(R,z)$ of the 2D tokamak cross-section, the total brightness, $b_i$, incident on a single detector, $i$, is given by 

\begin{equation}\label{eq:brightness}
    b_i = \int\hspace{-2mm}\int M_i(R,z) G(R,z) \,dR\,dz
\end{equation}

where $G(R,z)$ is the plasma emissivity distribution (in W/m$^3$) and $M_i(R,z)$ is a function that yields the relative contribution of $G(R,z)$ to $b_i$. This corresponds to evaluating the line integral of the plasma emissivity along the LOS corresponding to sensor $b_i$. 

\begin{figure*}[h!]
    \centering
        \includegraphics[scale=0.35, trim=0 0 550 0, clip]{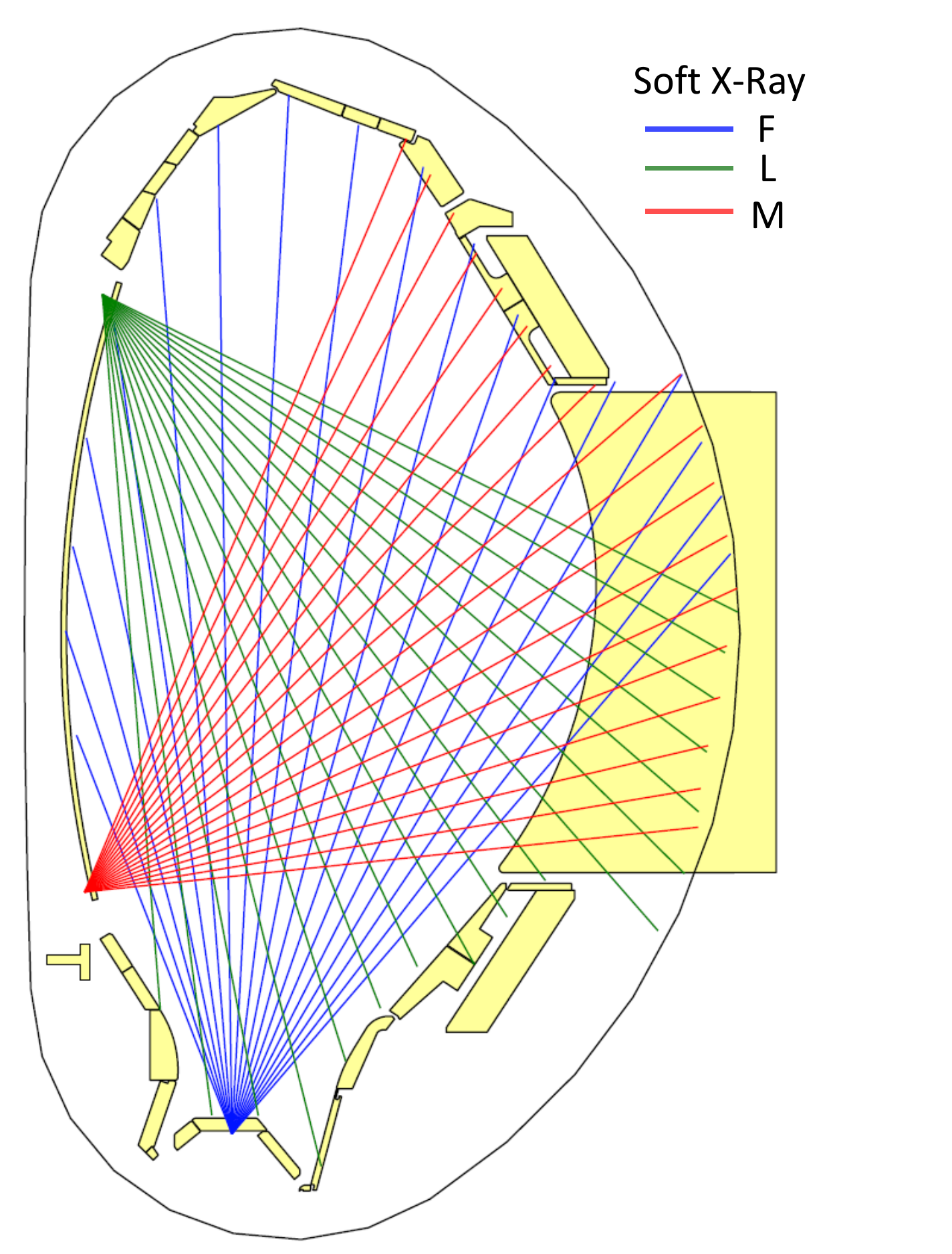}
        \caption{ASDEX Upgrade cross-section, and partial schema of the SXR measurement system, with three cameras (F, L and M) shown (plot obtained with \textit{diaggeom})}
        \label{fig:sxr}
\end{figure*}

By discretizing equation \ref{eq:brightness}, one obtains the plasma emissivity distribution at a finite number of positions (or pixels) along a tomographic reconstruction grid. In this case, the incident radiation on a single detector, assuming an associated noise, $\xi$, is\cite{odstrcil2016optimized}

\begin{equation}\label{eq:geo_mat_mapping}
    b_i = \sum_{j=1}^{n} M_{i,j} g_j + \xi_i \hspace{1cm }i \in 1, .., 208.
\end{equation}

From now on, we will denote the set of values $b_i$, that is, a set of $208$ line-integrated SXR measurements of the plasma emissivity taken at a certain point in time, as the plasma's tomographic projection in that instant. We denote equation \ref{eq:geo_mat_mapping} as the \textit{forward model} of the problem. Here, $n$ corresponds to the total number of pixels on a tomographic reconstruction grid, whereas $M_{i,j}$ is the discretization of the function $M(R,z)$ in equation \ref{eq:brightness}, mapping the relative contribution of pixel $j$ of that grid to measurement $i$ of the projection. The actual values of $M$ were pre-defined and contingent on the geometry and configuration of the sensors inside the machine, which can vary between different shot campaigns. Consequently, we denote $M$ as the \textit{geometric matrix}. The goal of a tomographic reconstruction algorithm is then to use the tomographic projection (i.e., the 208 measurements $b_i$) to find a suitable tomographic reconstruction $g$ that satisfies equation \ref{eq:geo_mat_mapping}. However, this problem is highly ill-posed\cite{murari2005development}, since small changes in projection space can translate into large changes in the tomographic reconstructions. It is also under-determined, meaning, the dimensionality of the reconstruction grid is much larger than that of the projection, ultimately resulting in an infinite number of solutions (reconstructions) that can fit the data. 

\subsection{Regularization-based Methods}

To solve the ill-posed problem, traditional tomographic algorithms use regularization techniques, usually based on assumptions regarding the smoothness of the plasma emissivity profile, that constrain the space of possible solutions. Such algorithms, however, are often computationally expensive and typically can only be used for post-experimental tomographic reconstruction, due to computational time constraints. In addition, the quality of the reconstructions is highly dependent on assumptions made about the data\cite{odstrcil2016optimized}.
Generally, those assumptions are encoded into the reconstructions through the use of Tikhonov regularization. In this case, computing the tomographic reconstruction of the plasma emissivity profile becomes a matter of finding a reconstruction $\hat{g}$ such that

\begin{equation}
    \hat{g} = \argmin_g (||Mg - b||^2 + \Lambda O(g))
\end{equation}{}

where $O(g)$ is a penalty term that encodes information about expected properties of the target plasma distribution, multiplied by a regularizing parameter $\Lambda$ that controls the regularization strength\cite{odstrcil2012modern}. There are several options for the choice of the regularization term $O$; typical choices are the Laplace operator, which favors smooth solutions, and minimum Fisher information\cite{anton1996x}, that favors solutions that are mostly flat in low-intensity regions, and peaked in high-intensity ones. 


\subsection{Deep Learning-based Methods}

Recent work has applied deep learning algorithms to the tomographic problem, namely by using de-convolutional neural networks to produce tomographic reconstructions taking measurement data as input\cite{matos2017deep}. This is achieved by training the networks on reconstructions that have been previously computed using standard tomographic algorithms. Generically, in a deep learning setting, a Deep Neural Network is \textit{trained} to learn a function mapping an input $x$ into its target output $y$\cite{goodfellow2016deep}; that is,


\begin{equation}
    y = y(x, \theta)
\end{equation}{}

where $\theta$ denotes the neural network's parameters, i.e. its weights and biases. The training process consists in finding an optimal value for $\theta$ that minimizes the mismatch between the network's outputs and their corresponding labels. 

In our setup, training a deep neural network to produce tomographic reconstructions would have required training it with measurements from the SXR diagnostic and pre-computed reconstructions, produced by other algorithms (namely, regularization-based ones). The expectation would then have been that the parameters $\theta$ computed during training would have converged to values such that if new, unseen data were fed into the network, it would be capable of \textit{generalizing} to outside of its training set. However, even assuming good generalization capacity of a neural network, it is at most as good as whatever data it has been trained on. In other words, should existing tomograms have had errors, a neural network would have learned to reproduce them. 



\subsection{Gaussian Process Tomography}

Another alternative is to use bayesian probability theory to produce tomographic reconstructions, by treating the underlying unknown plasma emissivity distribution as a \textit{Gaussian process}. Evaluating that process along a discrete set of points (the tomographic reconstruction grid) yields a multi-dimensional Gaussian distribution. 

By definition, in the Gaussian process framework, one assumes that multiple solutions for the tomographic reconstruction exist, in a Gaussian distribution of possible solutions. Treating the tomography problem with this framework allows using Bayesian formalism, which guarantees that the most likely solution for the tomographic reconstruction (i.e. the maximum \textit{a posteriori}, or MAP, estimate), subject to some assumptions about the underlying physical and data distributions, can be computed through Bayes's formula,

\begin{equation}\label{eq:bayesrule}
    P(A|B) = \frac{P(B|A)P(A)}{P(B)}.
\end{equation}{}

In the GPT setting, the terms in the formula are multivariate probability distributions, which are assumed to be Gaussian and are therefore specified by vectors of means (the individual means of each random variable) and a covariance matrix where each entry denotes the pair-wise covariance between those same variables. 

In Bayes's theorem, the term $P(A)$ is called the \textit{prior}. In GPT, by denoting the underlying plasma emissivity as $e$, the prior distribution $P(e) \sim \mathcal{N}(\mu_{pr}, \Sigma_{pr})$ encodes existing assumptions about the physical emission process, without observing any data (SXR measurements). Each random variable in the prior distribution is also Gaussian, and corresponds to the prior plasma emissivity $e$ at each point $x$ in the tomographic reconstruction grid. 

Here, the prior mean has a size equal to that of the tomographic reconstruction grid, $n$. Intuitively, the prior covariance matrix $\Sigma_{pr}$ encodes information about the expected smoothness of the plasma emissivity. The entries in the covariance matrix are computed for all pairs of points in the reconstruction grid through a prior covariance function. One covariance function generally used in Gaussian process regression is the squared exponential\cite{li2013bayesian}; in using this function, the prior covariance between a pair of points $x_1$ and $x_2$ in the tomographic reconstruction grid becomes

\begin{equation}\label{eq:squared_exp}
    \mathrm{cov}(x_1, x_2) = \theta_f^2 \exp \left(-\frac{(x_1 - x_2)^2}{2\theta_x^2}\right).
\end{equation}{}

The prior covariance is only dependent on the distance between points $x_1$ and $x_2$ and on $\theta = \{ \theta_f, \theta_x \} $, which are the model's \textit{hyperparameters}. The parameter $\theta_f$ controls the prior variance of the plasma emissivity at a given location in the reconstruction grid, whereas the parameter $\theta_x$, usually referred to as the \textit{length scale}, controls the extent to which different points in the reconstrucion grid are correlated. Models where the length scale is large yield high correlations even between grid points which are far apart, while smaller length scales yield covariance matrices where only points which are closer to each other are significantly correlated.


With these definitions, the prior becomes a probability distribution for the plasma emissivity, $e$, subject to the model's hyperparameters, i.e., $P(e|\theta)$, before any data, that is, a tomographic projection, has been observed. The prior can then be updated by multiplying it with the likelihood of the data $d$ (as per Bayes' theorem), yielding the posterior distribution, $P(e|d, \theta)$:

\begin{equation}
    P(e|d,\theta) = \frac{P(d|e,\theta) P(e|\theta)}{P(d|\theta)}
\end{equation}{}

The denominator in Bayes's theorem is known as the model \textit{evidence} or \textit{marginal likelihood}; if one is merely computing the posterior $P(e|d,\theta)$, it can be ignored, as it is just a normalizing constant. Interestingly, however, one can use this term to compare several different models (each with their own prior), and choose the one which best fits the data\cite{svensson2011non}. In this case, one assumes a \textit{hyper-prior}, from which different possible priors (individually specified by different hyperparameters) are sampled. The evidence can then be computed for different models -- a process that is referred to as \textit{marginalization} -- and the model with the highest evidence can be selected\cite{mackay1999comparison}. Calculating this requires an evaluation of the integral\cite{rasmussen2003gaussian}  

\begin{equation}
    P(d|\theta) = \int P(d|e, \theta) P(e|\theta) \,de .
\end{equation}{}

In practice, by assuming that the absolute value of the noise $\xi$ associated with the data (tomographic projection) also comes from a Gaussian distribution, the logarithm of the previous integral can be analytically calculated as\cite{rasmussen2003gaussian}

\begin{equation}\label{eq:log_marginal_likelihood}
    \fl log(P(d|\theta)) = - \frac{1}{2} \left( m \: log(2 \pi) + log|K+\Sigma_d| + (d-f_L)^T(K + \Sigma_d)^{-1} (d-f_L)\right)
\end{equation}

where $m$ is the number of SXR measurements in a tomographic projection, and $\Sigma_d$ is a diagonal matrix, whose non-zero entries are the individual variances, $\sigma^2$, of each measurement in the projection. Here, matrix $K$ is a linear transformation of the prior covariance $\Sigma_{pr}$ (imposed on the plasma emissivity) into measurement space, and is given by $K = M \Sigma_{pr} M^T$, where M is the geometric matrix defined in equation \ref{eq:geo_mat_mapping}. Similarly, $f_L$ is the mapping of the prior mean, $\mu_{pr}$,  (imposed in reconstruction space) onto measurement space, given by $f_L = M \cdot \mu_{pr}$.

The marginalization procedure is particularly useful because the trade-off between model complexity and data fit is automatic -- the model for which the evidence score is highest is always the simplest model that can explain the data, an embodiment of the Occam's Razor principle\cite{mackay1991bayesian}. 
In addition, the model evidence is also a function of the variance  $\sigma^2$ of the data (through matrix $\Sigma_d$), which means that it is possible to treat the expected projection error as an additional hyperparameter of the model to be tuned; this can be done, for example, by treating the data variance as a fraction of the measured value of SXR radiation in the tomographic projection. This means that, through the GPT framework, one can estimate not only the most likely model for the underlying plasma emissivity distribution, but also the most likely error values for the data itself. 

Once the most likely model is selected, and applying Bayes's formula, the posterior mean, $\mu_{post}$, and posterior covariance, $\Sigma_{post}$, as a function of the mean $\mu_{pr}$ and covariance $\Sigma_{pr}$ for that model, are respectively given by\cite{svensson2008current}

\begin{equation}\label{eq:post_mu}
    \mu_{post} = \mu_{pr} + \Sigma_{pr} M^{T} (K + \Sigma_d)^{-1} (d-f_L)
\end{equation}

and

\begin{equation}\label{eq:post_cov}
    \Sigma_{post} = \Sigma_{pr} - \Sigma_{pr} M^T (K + \Sigma_d)^{-1} M \Sigma_{pr}.
\end{equation}

By computing the posterior distribution $P(e|d, \theta) \sim \mathcal{N}(\mu_{post}, \Sigma_{post})$, one can then produce tomographic reconstructions either by sampling from $P(e|d,\theta)$, or simply by taking the mean of that distribution as the tomographic reconstruction (because the distribution is Gaussian, the mean corresponds to the MAP estimate). In addition, one can directly obtain uncertainties for the tomographic reconstruction from the diagonal values of the posterior covariance matrix, which correspond to the individual posterior variances of each pixel in the reconstruction grid.

The drawback of the marginalization procedure, however, is its potential computational complexity. First, the calculation of the evidence term involves a series of matrix multiplications and an inversion, which can be cumbersome particularly in our setting because of the dimensionality of the data, which generates very large matrices. Furthermore, the evidence must be computed for all models that are taken into consideration. When each model has several hyperparameters, the number of possible models to evaluate can become very large, which means that finding the optimal one can be time-consuming. For practical purposes, this limits the number of models which can be evaluated and thus, potentially limits the quality of the results. 

We therefore propose to bypass the need for analytical marginalization, by training a classifier (in this case, a CNN) to automatically choose the most likely model (out of several pre-defined ones) for the tomographic projection data collected by the ASDEX Upgrade SXR system.

This has potentially several advantages. On one hand, a GP model, while potentially having priors and posteriors with many dimensions, can be fully specified by its much smaller set of hyperparameters. In practice, this allows for parameterizing a distribution of high dimensionality with only a few variables. In the case of this work, this means that neural networks will learn to map tomographic projections to a lower-dimensional space (of dimensionality equal to the number of models under consideration). This should facilitate the network's learning process, allowing for easier generalization when compared with DL methods that attempt to map projections directly into a reconstruction space of larger dimensions. On the other hand, for potential real-time applications, this method potentially speeds up GPT, since it bypasses the marginalization procedure.

\section{Methods}
\label{sec:methods}

\subsection{Soft X-Ray Data}
\begin{figure*}[h!]
    \centering
        \includegraphics[scale=0.3, trim=40 0 40 50, clip]{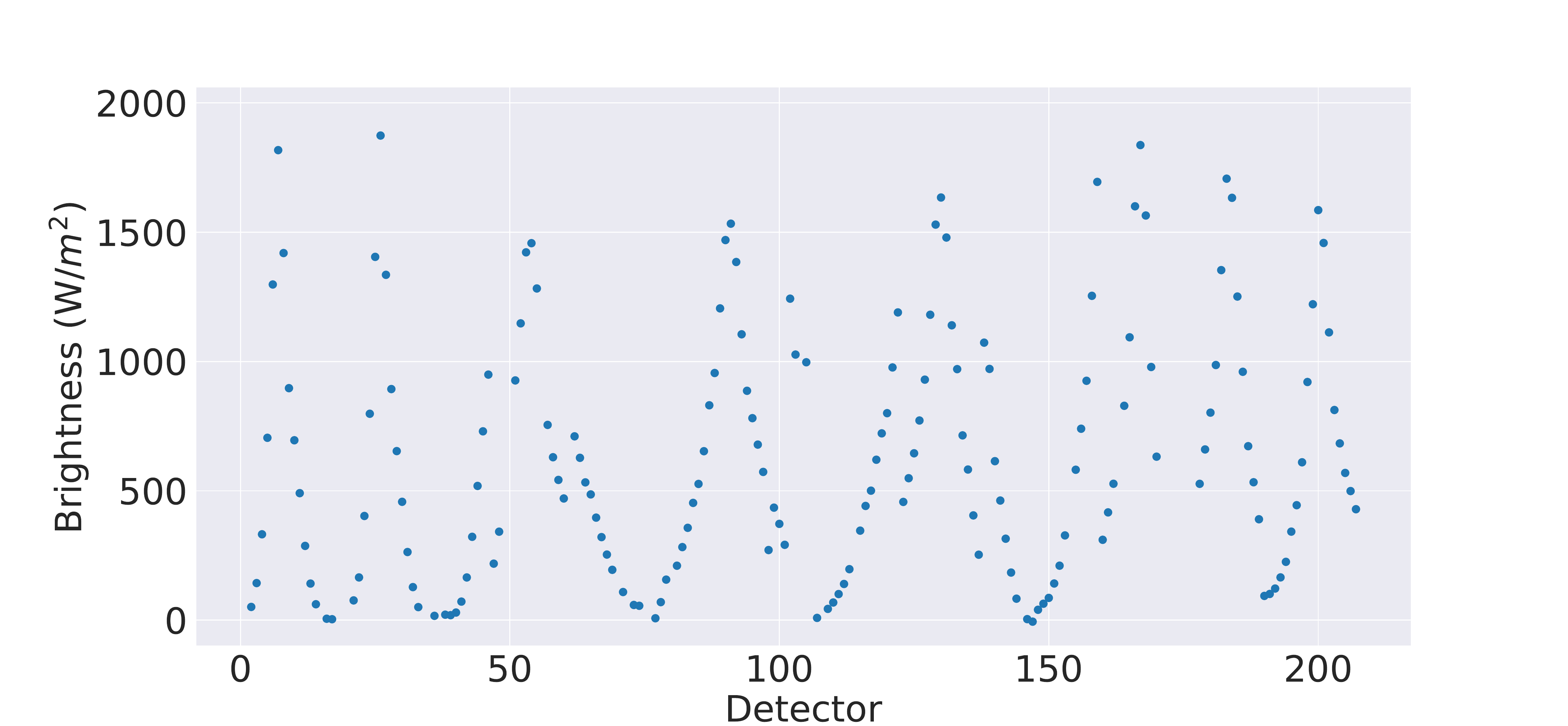}
        \caption{Sample tomographic projection (SXR measurements) from the ASDEX Upgrade tokamak, taken from shot $\#30294$ at $t=5,8691s$. Faulty measurements have been removed. }
        \label{fig:measurement}
\end{figure*}

For this work, we had at our disposal a collection of 112 ASDEX Upgrade shots, totalling $127 528$ data points (208-dimensional tomographic projections), with each dimension corresponding to a specific detector in the SXR system. The projections come from the down-sampled signal of the SXR diagnostic, at a sampling rate of $250 Hz$. The dataset also contains an error model, which assigns every measurement in every projection an estimated error value. In many cases the SXR detectors can be damaged and yield completely erroneous measurements, such as for example negative brightness; in these cases, the corresponding error values in the existing error model are infinite. A sample projection can be seen in Figure \ref{fig:measurement}. 

We also possessed a geometric matrix M that maps the relative contribution of each pixel in a $60\times40 (2400)$-dimensional tomographic reconstruction grid to each of the 208 SXR measurements in a projection. Each pixel in the grid has a pair of poloidal coordinates $(R,z)$, based on the poloidal dimensions of ASDEX Upgrade; the tomographic reconstruction is computed on this grid. The geometric matrix itself was computed based on the physical layout of the SXR sensors in the ASDEX Upgrade vessel, and holds for all shots in our dataset. 

\subsection{Dataset Generation}
Before training the neural network classifier, we generated its training and validation dataset by individually computing, for all the measured tomographic projections, the most likely model for the plasma emissivity distribution that generated those same projections. To that end, the first task was to define a hyper-prior, i.e., a prior for the models' hyperparameters; then, different models (individually specified by their specific set of hyperparameters) were compared based on their evidence. 

As is typical in Gaussian process regression tasks \cite{rasmussen2003gaussian}, we defined all prior means, $\mu_{pr}$, as vectors of zeros, of size $2400$ (the size of the reconstruction grid). We computed the prior covariance matrices $\Sigma_{pr}$ using a squared exponential function as defined in equation \ref{eq:squared_exp}; in essence, this covariance function encodes our belief that correlations between pixels on the tomographic reconstruction grid will decay exponentially as the distance between those points increases. The covariance function has only two parameters: $\theta_f$, the individual variance of single pixels, and $\theta_x$, the length scale which controls the extent of the correlation between pixels in the reconstruction grid. Given these parameters, the covariance function computed the covariance between two points $x_1$ and $x_2$ as a function of the distance between them. We computed the distance between two points in the reconstruction grid (expressed in terms of their poloidal coordinates) using the Euclidean definition, i.e., $d(x_1, x_2) = \sqrt{(R_1-R_2)^2 + (z_1 - z_2)^2}$. 

We use the information in the data's existing error model to discard faulty measurements -- that is, for a given projection, we discard measurements whose associated error is infinite. In practice, this means that, when carrying out the marginalization process, and computing the MAP estimate for the plasma emissivity, some of the $208$ measurements of each projection were not used. In addition, different projections have different damaged sensors. We treat the remaining measurements in a projection as the mean values of a multivariate gaussian distribution, where the variables are independent from each other. The variances, $\sigma^2$, of that distribution correspond to the entries in the diagonal of matrix $\Sigma_d$ of equation \ref{eq:log_marginal_likelihood}, and represent the uncertainties in the measurements, derived from their noise. In a gaussian distribution, approximately $99.73\%$ of the data falls within $3$ standard deviations ($\sigma$) of the mean (i.e. the  3-$\sigma$ rule); we use this rule to assume that, if $\xi_i$ is the absolute value of the error of measurement $i$, then $3\sigma_i = \xi_i$. 


One possibility would then have been to compute the variance of each measurement in a projection based on the values of $\xi$ in the existing error model. In this work, however, we instead computed the value of $\xi$ as a fraction of the noisy measurements themselves, by treating that fraction as an additional hyperparameter, $\theta_{err}$, to be optimized through Bayesian marginalization. We note, however, that we assumed this value to be global, i.e., the same for all measurements in a projection; therefore, for a single SXR measurement $m_i$, $\xi_i = \theta_{err} m_i$. Given our earlier assumption using the $3\sigma$-rule, the absolute values of the variances, $\sigma^2$, of each measurement $i$ could then be calculated as $\sigma^2 = \frac{1}{9} \theta_{err}^2 m_i^2 $; treating $\theta_{err}$ as a scaling factor is the simplest assumption that can be made about the data errors, apart from using the existing error model. 

Formalizing, we iteratively computed, for each individual data point $d$ (i.e. projection), the highest-evidence model for the plasma emissivity and data distributions that generated that projection, that is, through equation \ref{eq:log_marginal_likelihood} we looked for $ \Hat{\theta} = (\Hat{\theta_f}, \Hat{\theta_x}, \Hat{\theta_{err}})$ such that

$$\Hat{\theta} = \argmax_{\theta} \log p(d|\theta).$$

We searched for the ideal hyperparameters in a grid by assuming a uniform hyper-prior, and computed the model evidences at several discrete positions in the hyper-prior space. The question was then, what positions in the hyper-prior space should one evaluate the models' evidences on? This required taking several factors into account. 

The first requirement was the expected nature of the plasma emission process itself. A previous analysis of the measurement data, and of existing tomographic reconstructions from ASDEX Upgrade\cite{odstrcil2016optimized}, showed that the plasma emissivity has a wide dynamic range for different regions of the plasma, with emissivity in the plasma core being up to 3 orders of magnitude higher than in the pedestal. Likewise, in some periods of some shots, the maximum radiation value in the reconstruction grid was in the order of magnitude of $10^2 W/m^{-3}$, while in other phases, it could be as large as $10^5 W/m^{-3}$. Thus, we considered this range in emissivities a good region to explore possible values for the hyperparameter $\theta_f$. In addition, ASDEX Upgrade has a minor radius $a=0.5m$ (horizontally) and $b=0.8m$ (vertically)\cite{lechte2017x}; given this and the size of our reconstruction grid, we assumed that a good region of the hyper-prior in which to evaluate the evidence for certain values of $\theta_x$ ranged, in the limit, from $0$ (no correlation at all between pixels) to $1.6$. For the hyparameter $\theta_{err}$ we assumed that, in the limit, it could range from $0$ (no noise in the tomographic projections) to $1$ (all of the measured brightness corresponded to noise). 

The second requirement related to the training process for neural networks. For this work, we wanted to train a neural network to perform a classification task -- to learn to map measurements to the the most likely model. Typically, in a machine learning classification setting, care should be taken such that training samples fed to a network are reasonably balanced with respect to their different classes; that is, a good training practice is that one class not be too over-represented in the data when compared to others. 

\begin{figure*}[h!]
   \centering
        \includegraphics[scale=0.6, trim=0 0 0 0, clip]{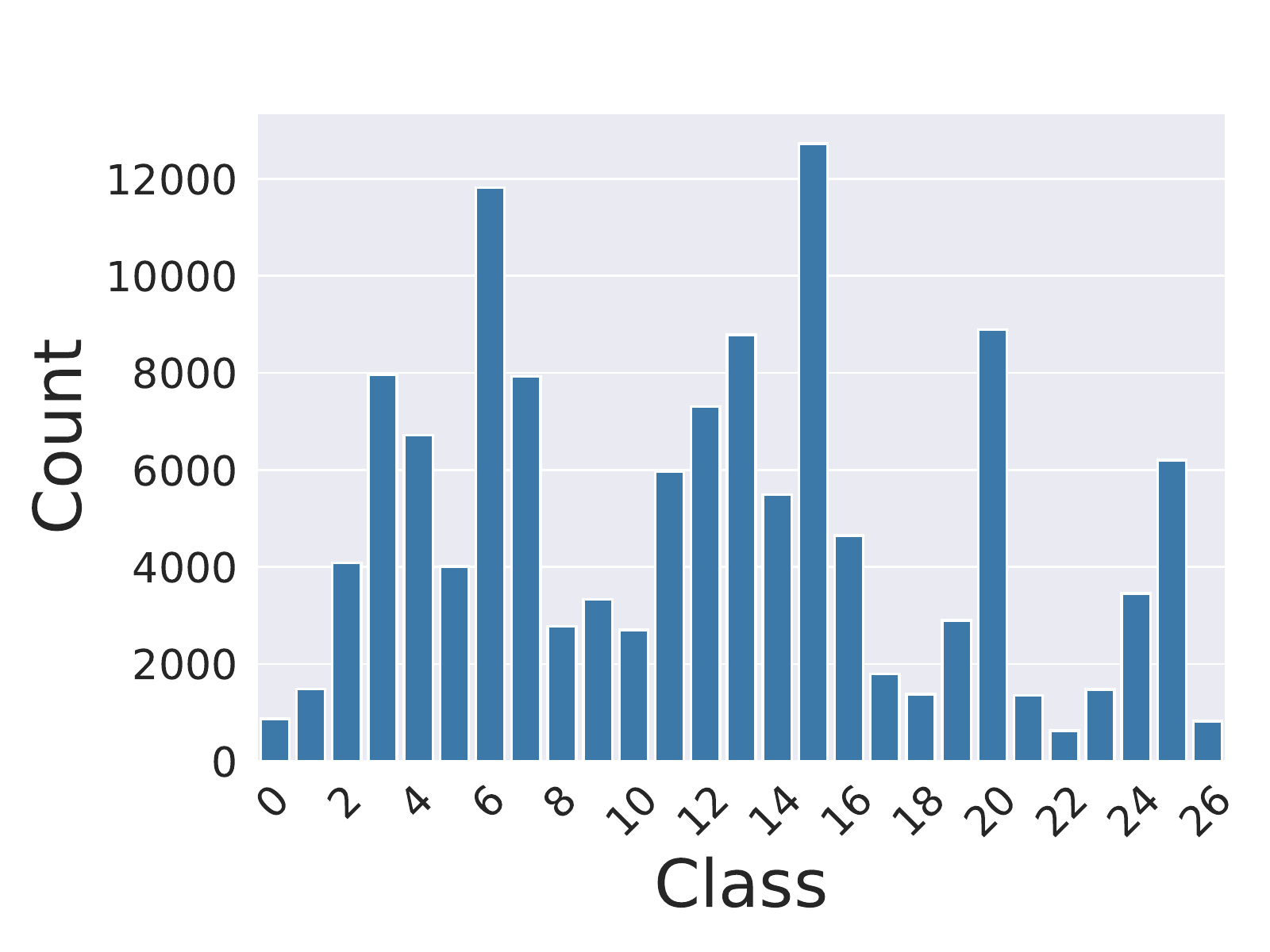}
        \caption{Number of data samples (from all available shots) mapping to each class (set of hyperparameter values), after the marginalization procedure.}
        \label{fig:heatmap}
\end{figure*}

In practice, considering these requirements, we performed several evaluations through trial and error of the hyper-prior at different positions; we settled on a $3\times3\times3$ grid, where the points correspond to $\theta_f = \{500, 1000, 2500\}$, $\theta_x=\{.15, .175, .2\}$ and $\theta_{err}=\{.5, .75., 1\}$, which corresponds to 27 GP models. Performing the Bayesian model selection procedure on all projections in our dataset using the models parameterized by these values of $(\theta_f, \theta_x, \theta_{err})$ yielded a relative balance in terms of the amount of data samples mapping to each of the 27 possible classes (points on the hyperparameter grid); this can be seen in Figure \ref{fig:heatmap}, which shows the number of points mapping to each class. Computing the evidence for different models for all tomographic projections took a total of $48$h. 
This dataset -- i.e., the mappings between tomographic projections and the class to which their highest-evidence model belongs (out of 27 possible ones) -- was then used to train and test the neural network classifier.

\subsection{Deep Learning Model}

Several possibilities exist when it comes to modelling deep neural network architectures. For our purposes (the learning of the Bayesian model selection procedure) we opted to use a CNN. CNNs are widely used for signal processing tasks, due to their ability to efficiently detect spatial correlations in data, which is what we expected to find in our SXR measurements. The model we used is, with regards to its architecture, inspired by the VGG network for classification of images\cite{simonyan2014very}. We designed the model using the Keras framework for deep learning\cite{chollet2015keras}.

\begin{figure*}[h!]
   \centering
        \includegraphics[scale=0.9, trim=0 0 0 0, clip]{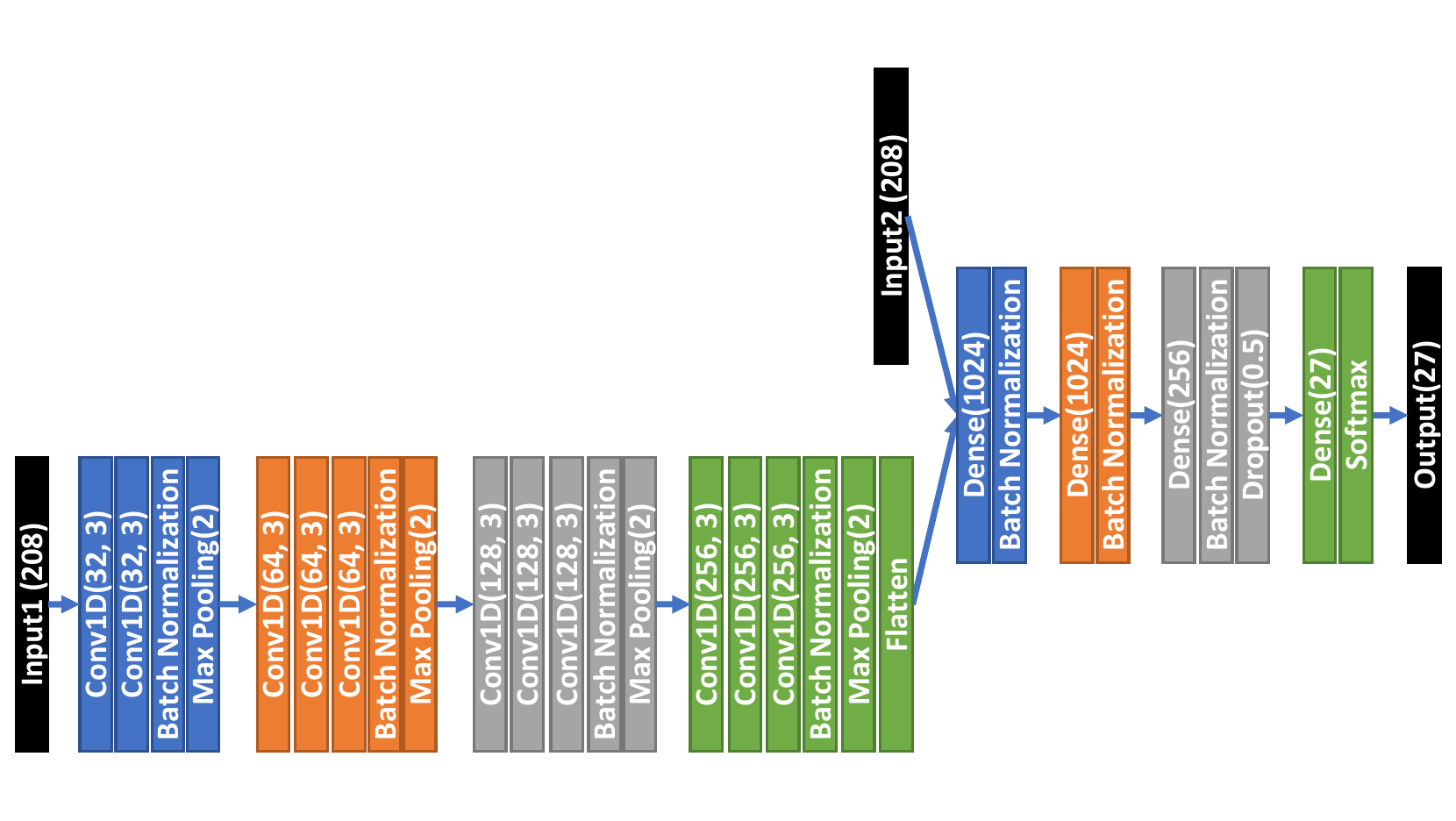}
        \caption{Schematic of the deep learning model used for this work. }
        \label{fig:cnn_arch}
\end{figure*}

The network itself receives two inputs: a tomographic projection (208 SXR measurements), and a corresponding mask of ones and zeros (taken from the existing error model in our dataset), that gives information regarding which measurements in the projection are assumed to be faulty. The network uses a series of convolutional layers followed by max pooling layers to process high-level features in the measurement data. The output of those layers is then combined with the information in the error mask and processed in the last layers of the network, which are standard fully-connected layers. We also used batch normalization\cite{ioffe2015batch} layers to speed up training, and dropout in the final layer\cite{li2019understanding} to increase the network's capacity for generalization outside of its training set. We used the rectified linear unit (ReLU) activation function throughout the entire network apart from the last layer, which uses a softmax function, because we modelled the network output as probabilities over $27$ possible classes, which must add up to 1. For the same reason, we used categorical cross-entropy as loss function. We used the Adam optimizer\cite{kingma2014adam}, and left all optimizer hyperparameters at their default values. Figure \ref{fig:cnn_arch} shows a schematic of the neural network architecture.

\section{Results}
\label{sec:results}

We here performed two separate assessments. First, we evaluated the accuracy of the neural network's fit of the individual projections to their highest-evidence models. Then, based on the highest-probability class determined (by the network) for each data point, we computed the corresponding MAP estimate of the tomographic profile, and measured the quality of those reconstructions by projecting them back into measurement space (through the forward model in equation \ref{eq:geo_mat_mapping}), obtaining their \textit{back-projections}. We then measured the percentage error of those back-projections when compared to the original tomographic projections.

\subsection{Neural Network}

To increase the robustness of our methods, we opted to train an ensemble of neural networks (of equal architectures), using the k-fold cross-validation strategy\cite{james2013an}. K-fold cross validation is useful to determine whether the choice of the train/test split has biased whatever results have been obtained, or whether the results can be assumed to hold independently of the data split. We opted to divide our data into k$=10$ folds -- that is, we trained 10 networks with different overlapping splits of train data, and tested them on non-overlapping validation splits. We trained the networks for 50 epochs, and ran them on an NVIDIA Quadro RTX 5000 Graphics Processing Unit. The total training time for the whole ensemble was 1h, while total prediction time for the validation data was $41,62$s. 

As the networks performed a 27-way classification, we used top-k categorical accuracy as a metric for network classification quality. We now follow with a brief explanation of this metric. 

Each data point $x$ (corresponding to a tomographic projection) in our dataset was assigned a label, $y_{label}$, denoting for which of the 27 model classes the evidence was highest. A classifier learns, through the training process, to compute the probability of that point belonging to a certain class $P(C(x) = c)$, where $c$ can take one out of 27 possible values; we denote the vector containing the probabilities of belonging to each of those classes $y_{pred}$. We further define $y_{pred_k}$ as the $k-$th most likely class given by a classifier for $x$; for example, for $y_{pred_{1}}$, one would get

$$y_{pred_{1}} = \argmax_c P(C(x) = c) = \argmax_c y_{pred} $$

whereas for $c_{27}$ one would have

$$y_{pred_{27}} = \argmin_c P(C(x) = c) = \argmin_c y_{pred}.$$

Based on this, the top-k accuracy metric then calculates for each data point:

$$
acc_k(x) =\cases{ 1& if $y_{label} \subset \{y_{pred_{1}}, ... y_{pred_{k}}\}.$ \\ 0& otherwise.\\} 
$$

\begin{figure*}[h!]
   \centering
        \includegraphics[scale=0.5, trim=0 0 0 0, clip]{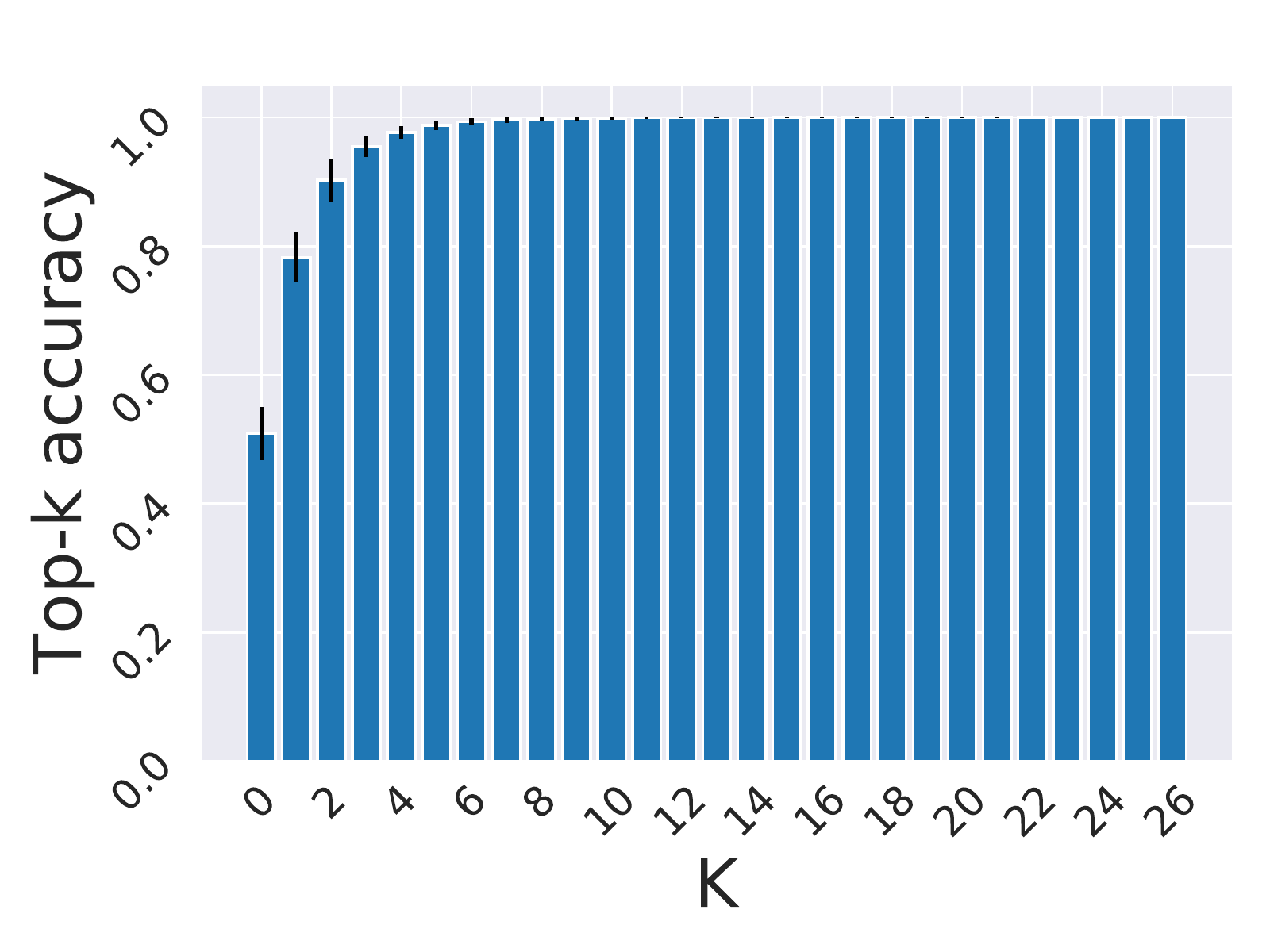}
        \caption{Top-k accuracy (up to k=$27$) for validation data. The blue bars indicate the mean accuracy across the ensemble of 10 networks, while the smaller black bars indicate the accuracy's standard deviation across the ensemble (for each k)}
        \label{fig:topk_class}
\end{figure*}

We then computed the categorical accuracy metric on the validation data for the 27 different values of k. Because we opted for a cross-validation train and test strategy (with an ensemble of 10 classifiers), we show the mean value and standard deviation of the top-k accuracy across all members of the ensemble. The results of the metric can be seen in Figure \ref{fig:topk_class} (up to k=27) and Table \ref{tab:top5results} (up to k=5). In Table \ref{tab:top5results} we show the results only up to k=5 for ease of comprehension. 

\begin{table}[h!]
    \begin{center}

        \begin{tabular}{c|c|c|c|c|c|}
        \cline{2-6}
         & \multicolumn{5}{c|}{K} \\ \cline{2-6} 
         & 1 & 2 & 3 & 4 & 5 \\ \hline
        \multicolumn{1}{|c|}{mean} & 0.509 &  0.783  & 0.903  & 0.955  & 0.977 \\ \hline
        \multicolumn{1}{|c|}{st. dev.} & 0.041   & 0.038 & 0.033 & 0.016  &  0.01 \\ \hline
        \end{tabular}
        \caption{Accuracy mean and standard deviation across ensemble of 10 neural networks, for validation data, up to top-5 accuracy.}
    \end{center}
\end{table}
\label{tab:top5results}

\subsection{Sample Reconstructions}

In addition to evaluating the neural network's capacity for classification purposes, we also produced and evaluated tomographic reconstructions. To that end we took, for each data point in the validation dataset, the most likely class prediction given by the neural network; this class prediction maps to one of the models we have previously defined. We then computed, based on the class prediction and on equations \ref{eq:post_mu} and \ref{eq:post_cov}, the posterior mean and covariance for each data point. We took the posterior means (i.e. the MAP estimates) as the tomographic reconstructions of the plasma emissivity - each mean was a 2400-dimensional vector, where each entry $\mu_j$ denotes the most likely value for the plasma emissivity in a point $j$ in the reconstruction grid. The posterior covariances allowed us to determine the error of the tomographic reconstruction, by taking the diagonal of the covariance matrix, which corresponds to the individual variance $\sigma_{post}^2$ of each pixel in the reconstruction grid; we converted the value of that variance into a percentage error by once again taking advantage of the $3-\sigma$ rule, and computing said percentage, for pixel $j$, as

$$\%_{err_j} = 3 \frac{\sqrt{\sigma_{post_j}^2}}{\mu_j} \times 100\%$$

\begin{figure*}[h!]
   \centering
        \includegraphics[scale=1.0, trim=200 100 200 140, clip]{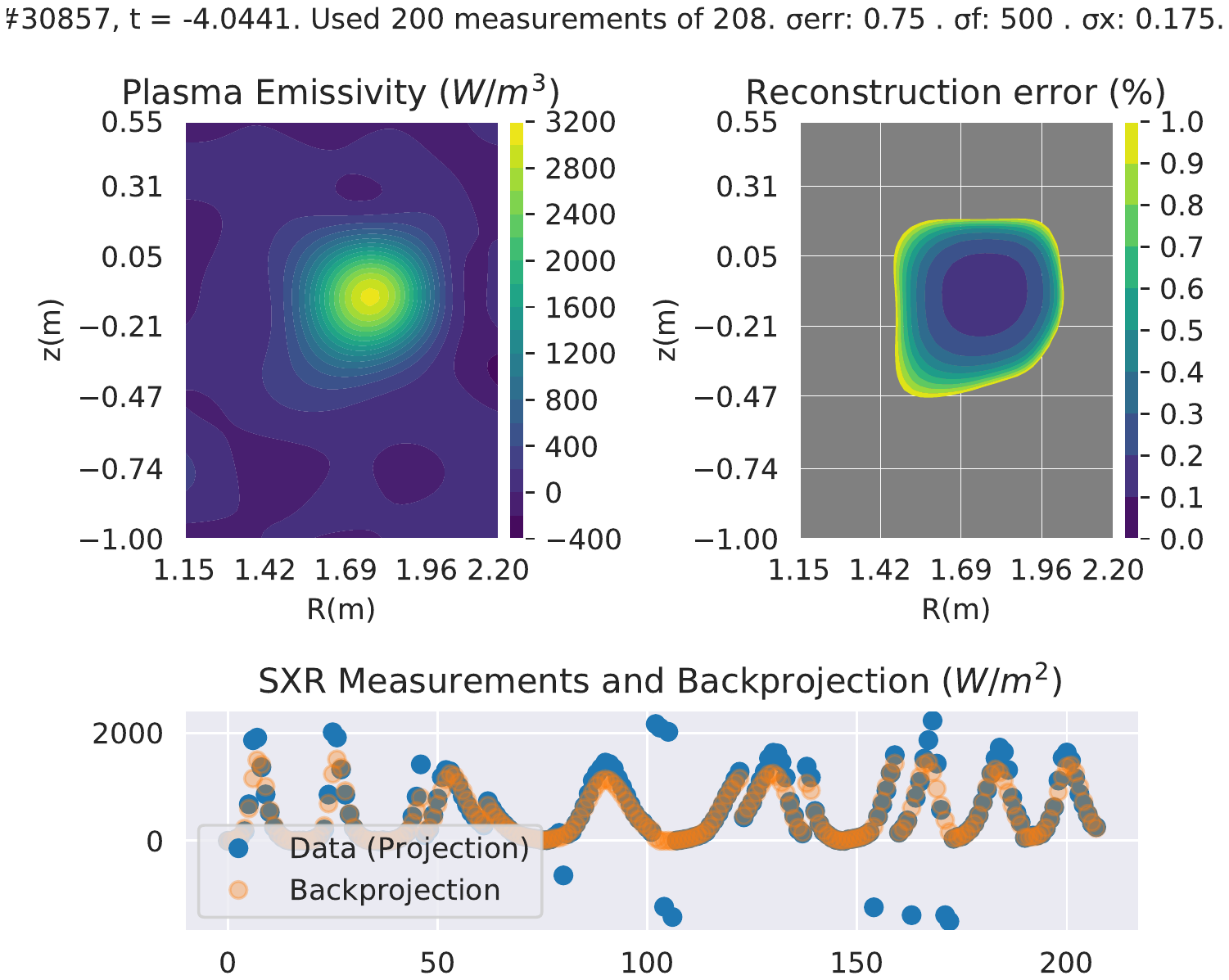}
        \caption{Sample tomographic reconstruction and error, and comparison between the SXR measurement and the back-projected reconstruction, for ASDEX Upgrade shot \#30857, t=$4,0441s$. The determined model hyperparameters by the classifier were $\theta_{err}=0,75, \theta_f=500, \theta_x=0,175$; 200 measurements (out of 208) were used for this reconstruction. }
        \label{fig:rec1}
\end{figure*}

\begin{figure*}[h!]
   \centering
        \includegraphics[scale=1.0, trim=200 100 200 140, clip]{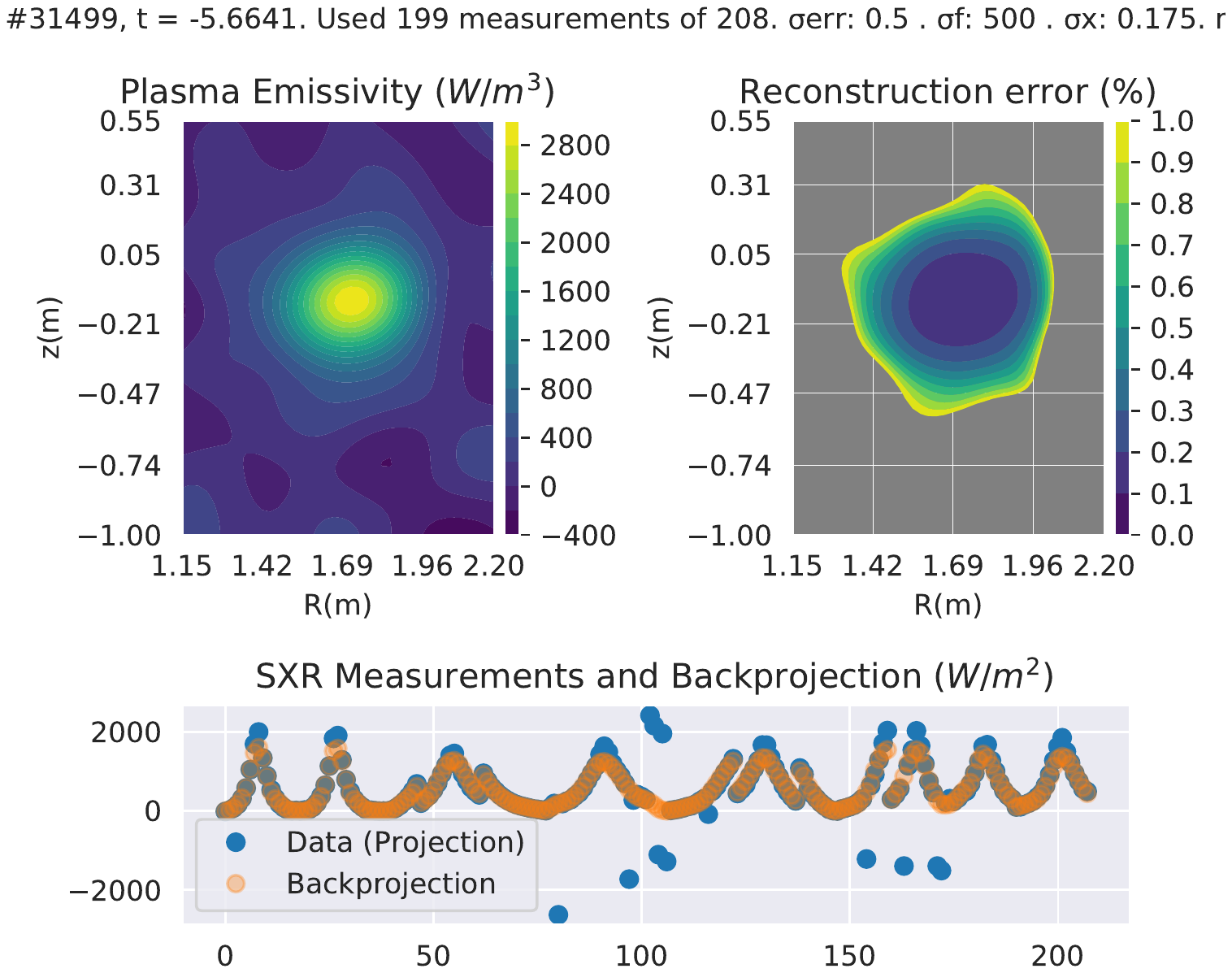}
        \caption{Sample tomographic reconstruction and error, and comparison between the SXR measurement and the back-projected reconstruction, from ASDEX Upgrade shot \#31499, t$=5,6641$. The determined model hyperparameters by the classifier were $\theta_{err}=0,5, \theta_f=500, \theta_x=0,175$. 199 measurements (out of 208) were used for this reconstruction.}
        \label{fig:rec2}
\end{figure*}

Two sample results can be seen in figures \ref{fig:rec1} and \ref{fig:rec2}. For the reconstruction error, we show only points where the percentage error was found to be below $100\%$. 

\clearpage
\subsection{Regression Error}

\begin{figure}[h!]
    \centering
    \begin{subfigure}{0.5\textwidth}
       \centering
        \includegraphics[scale=0.45, trim=0 20 0 0, clip]{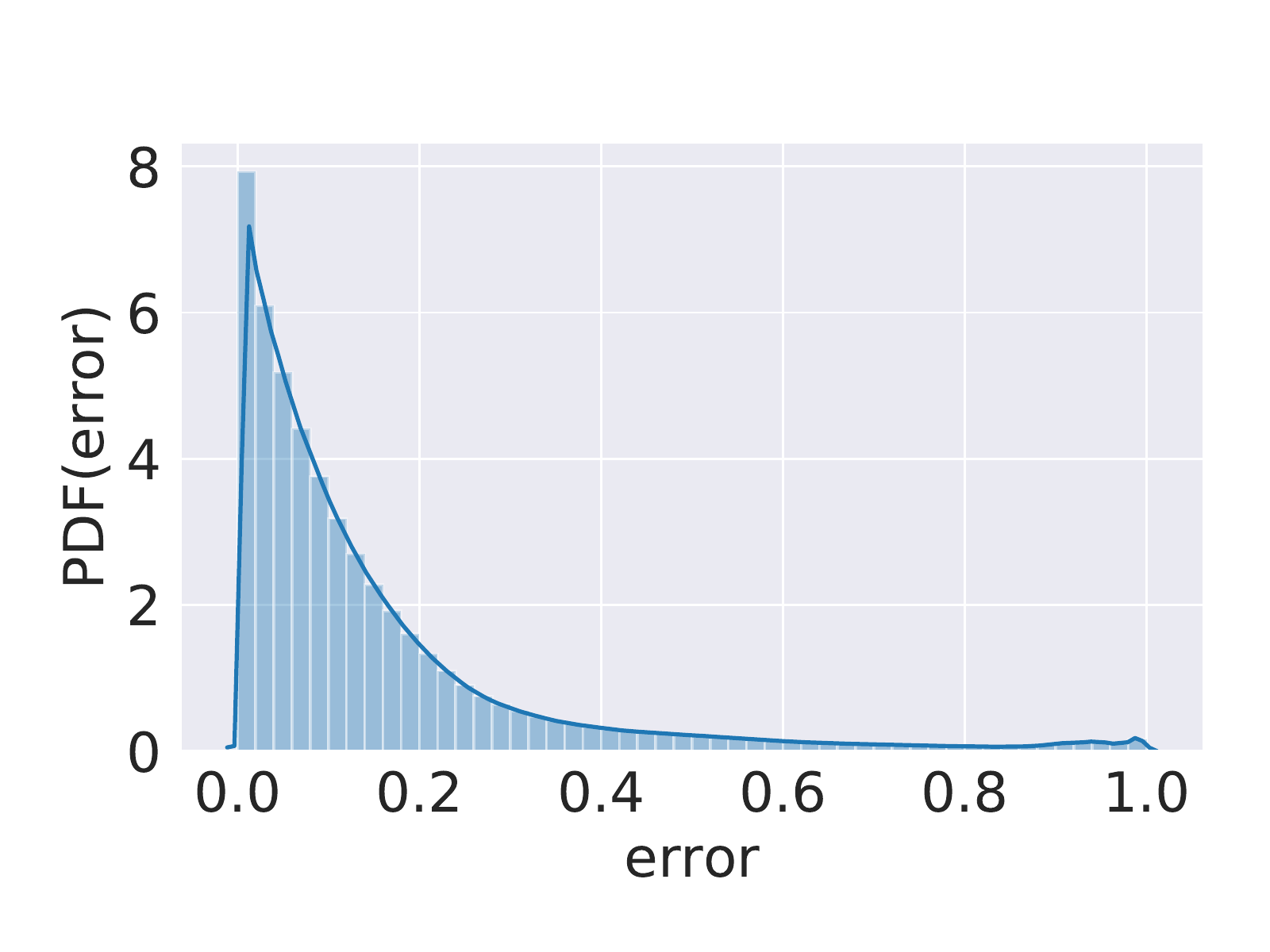}
        \caption{MAE values}
        \label{subfig:err_histogram}
    \end{subfigure}%
    ~
    \begin{subfigure}{0.45\textwidth}
       \centering
        \includegraphics[scale=0.45, trim=0 20 0 0, clip]{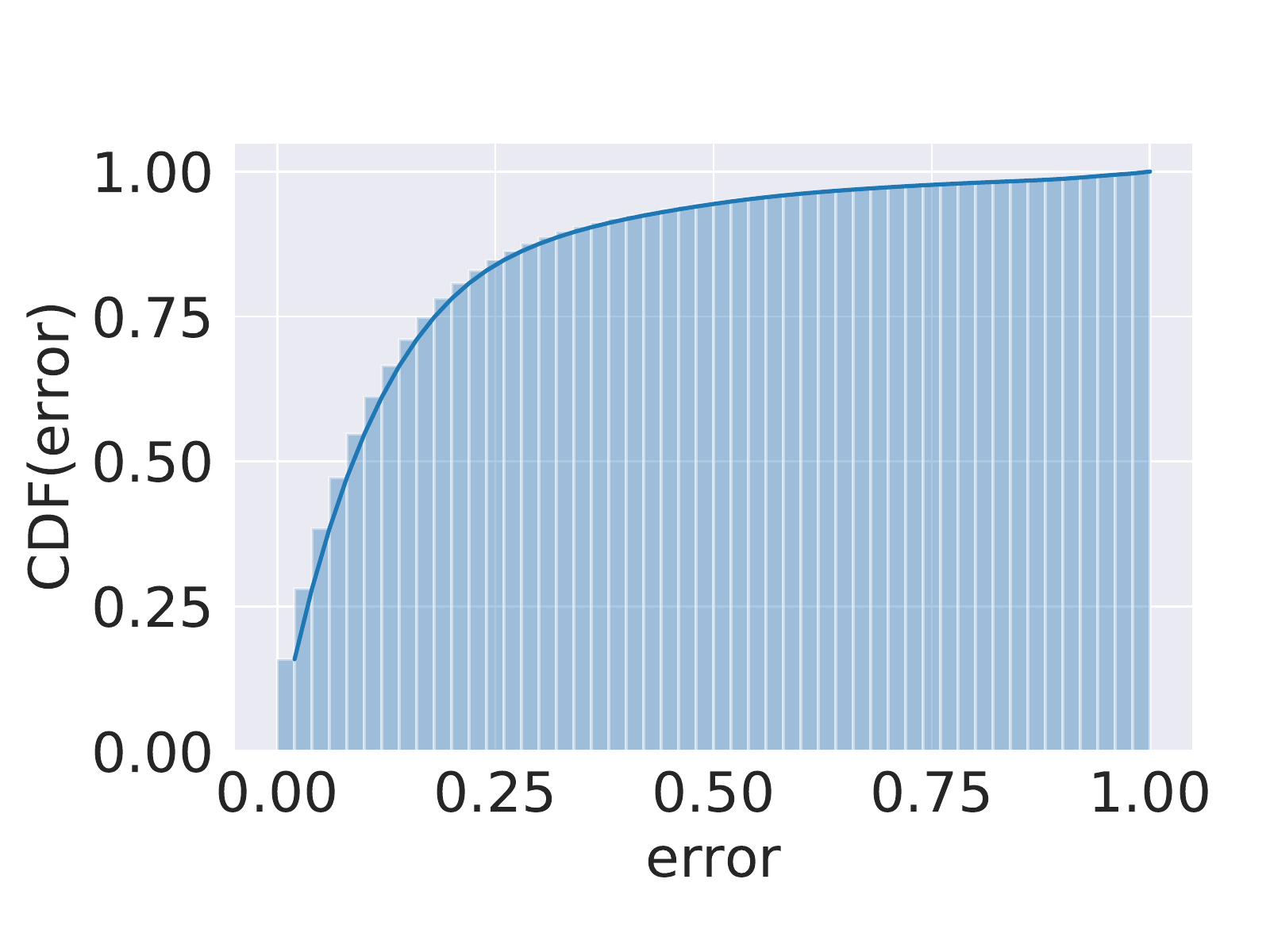}
        \caption{Cumulative MAE values}
        \label{subfig:err_histogram_cum}
    \end{subfigure}%
    \caption{Cumulative distribution of the error values of the tomographic reconstructions' back-projections into measurement space. $54,4\%$ of the individual back-projected measurements have a relative error lower than $10\%$; $93.83\%$ have an error lower than $50\%$; and $99,38\%$ have an error lower than $100\%$.}
\end{figure}

To evaluate the quality of the tomographic reconstructions, we performed, for each reconstruction, a pass through the forward model defined in equation \ref{eq:geo_mat_mapping} to obtain the corresponding back-projection, i.e., the projection of the reconstruction back into measurement space. Performing this check allowed to see how well the obtained MAP estimate actually fits the original SXR data; to do so, we computed the percentage error between the back-projection and the original tomographic projection, and did this for every measurement in every data point. The histograms in Figures \ref{subfig:err_histogram} and \ref{subfig:err_histogram_cum} show the results of this evaluation, up to an error of $100\%$, a threshold which covers $99,38\%$ of the validation data. Note that the error was computed by comparing the back-projection only with valid measurements (i.e. which are not labeled by the existing data model as having infinite error). On average, $91.25 \%$ of the 208 measurements in each data point were used to compute the tomographic reconstructions.

\subsection{Discussion}

An analysis of Table \ref{tab:top5results} and Figure \ref{fig:topk_class} shows that the ensemble of 10 neural networks achieves very good results on the classification task, with a mean top-5 accuracy score of $0.976$ (out of a maximum score of $1$) for validation data. Furthermore, the standard deviation of the accuracy score demonstrates consistently low values, indicating that the choice of train/test split for our data did not significantly bias the achieved results; all neural networks in the ensemble behave similarly, even if tested on different data. 

Taking the highest-probability model hyperparameters computed by the classifiers for the entire validation set, and computing the corresponding MAP estimates of the tomographic reconstructions, we similarly see good results; the reconstructions' back-projections mostly show very good agreement with the original tomographic projections, with more than $90\%$ of the back-projected data having an error lower than $50\%$.

\section{Conclusions}
\label{sec:conclusions}

Gaussian process tomography makes it possible to obtain the most likely estimate for an unknown, potentially infinite-dimensional, quantity, given some assumptions about the underlying physical distribution and about the data generated by that distribution. The tomography problem, based on SXR measurement data from the ASDEX Upgrade tokamak, lends itself to investigation under this framework. If one assumes a fixed model for the behavior of the underlying physical distribution (i.e. the plasma emissivity) and for the data, for example by specifying the length-scales involved in the emission process and the expected fraction of noise in the measurements, GPT inversion techniques readily yield the corresponding maximum \textit{a posteriori} estimate of the plasma SXR emissivity in the two-dimensional tokamak cross-section. 

Nevertheless, this raises the issue of what models one would like to assume in the first place. Through the Bayesian Occam's Razor principle, GPT answers this question by computing the evidence for different possible models, out of which the one with the highest score can then be selected. This can be useful if one wishes to test different assumptions regarding the data distribution: for example, what fraction of noise can be expected in the observations (measurements)? However, in a setting such as SXR tomography with ASDEX Upgrade data, this task can become cumbersome due to the dimensionality of the tomographic projections. This is further compounded when the number of models under evaluation is large. 

For that reason, we developed a novel method for automatic selection of the best model (out of 27 pre-defined ones) for the plasma SXR emissivity distribution and the corresponding data, for measurements from the ASDEX Upgrade tokamak. The individual models had different assumptions regarding the noise level in the collected data, the correlations between variables in the tomographic reconstruction grid, and the individual variances of those same variables. The method then consisted in training a convolutional neural network to perform the bayesian model selection (marginalization) procedure, and bypass the need to perform that task analytically. Our results show that the neural network achieved good classification results when compared to the analytical bayesian marginalization step, with top-5 accuracy (out of 27 possible classes) reaching a value of 0.976 (out of a maximum of 1). Furthermore, while the marginalization procedure across the entire dataset (of $127 528$ tomographic projections), through analytical methods, took approximately $48$h, the same computation, performed by the neural network, took only $43$s. Thus, the neural network approach can be particularly useful for high-dimensional data settings such as ours, as well as problems where the number of models under consideration is large, which would otherwise render the model comparison problem too slow through analytical methods. This can be particularly useful for settings where not only real-time inversion of tomographic profiles, but also real-time comparison of different models for certain physical distributions is a necessity.

\section*{Acknowledgements}
\small
\noindent This work has been carried out within the framework of the EUROfusion Consortium and has received funding from the Euratom research and training programme 2014-2018 under grant agreement No 633053. The views and opinions expressed herein do not necessarily reflect those of the European Commission.

\clearpage
\section*{References}
\bibliography{iopart-num}

\end{document}